%Paper: chao-dyn/9303003
%From: ashurs@california.sandia.gov (ashurst william t)
%Date: Wed, 10 Mar 93 15:57:00 -0800
%Date (revised): Wed, 10 Mar 93 16:38:30 -0800

\magnification = 1200
\baselineskip 18pt

\centerline{ to appear in Combustion Science \& Technology, 1993 }
\vskip 12pt
\centerline{ \bf Flame Propagation Through Swirling Eddys, A Recursive Pattern
}

\vskip 12pt

\centerline{ Wm. T. Ashurst }

\centerline{ Combustion Research Facility }

\centerline{ Sandia National Laboratories }

\centerline{ Livermore, California 94551-0969 }

\vskip 12pt

{\it Abstract -- Computed flame motion through and between swirling eddys
exhibits a maximum advancement rate which is related to the time duration of
flame motion between eddys.  This eddy spatial structure effect upon the
apparent turbulent flame speed appears to be similar to the square-root
dependence observed in wrinkled flamelet data. The rate-limiting behavior at
one eddy length-scale can be removed by inclusion of smaller eddys which reside
between the larger eddys.  This large-eddy, small-eddy concept yields a
recursion relation and repeated functional iteration can be done to approximate
a desired flame speed relation.  As an example, an iteration to produce $S_T
\ln
S_T = u'$ is given for the range of $u'$ observed in liquid flames.  Currently,
the iteration process is a post-diction of flame speed, but if a universality
can be developed, then a predictive theory of turbulent flame propagation might
be achieved. }

\vskip 24pt

\centerline{ Prediction of Premixed Turbulent Flame Speed}

Determination of the propagation rate of a flame through a turbulent gas with
premixed fuel and air is a very practical problem with many attempts since the
work by Damk\"ohler in 1940.  He suggested that if the turbulent motion did not
change the local flame speed from its laminar value $S_L$, then the effective
turbulent flame speed was proportional to the total flame area divided by the
cross-sectional flow area: $S_T/S_L = A_T/A$.  But, determination of the total
flame surface area by direct measurement or calculation has not been possible.
A year after Damk\"ohler, Kolmogorov proposed the inviscid energy cascade in
turbulent motion and so, the idea that motions at different wavelengths have a
certain power relation has been applied to the flame propagation problem.
Application of Kolmogorov's 1941 theory to the premixed flamelet regime leads
to $S_T \sim u'$, where the energy dissipation rate is $u'^3/L$ and $u'$ is
the rms velocity with $L$ the velocity integral length scale (see Clavin \&
Siggia, 1991).  In this model the turbulent propagation rate does not depend
upon the laminar flame speed, nor is there an upper limit on $S_T$ as $u'$
increases.  Also there is no explicit description of the spatial structure of
the tubulent flow.

Peters introduced the corrugated flamelet regime with the Gibson length scale
for conditions where the flame thickness is smaller than the smallest turbulent
scale and the laminar flame velocity is small compared to the velocity
fluctuations (1988).  The Gibson length corresponds to the smallest wavelength
of the flame wrinkles, because turbulent eddys which are smaller than $L_G$
have
rotational velocities smaller than $S_L$ and so can not distort the flame
surface.  Kerstein (1988a) suggested that each active eddy length scale between
$L_G$ and $L$ will burn out in the time required for one turn over time of that
particular eddy, but the rate of propagation within that eddy is enhanced by
the smaller eddys.  If the fractal dimension is 7/3, then this consumption of
an eddy in one turn over time will occur for all eddy sizes over the fractal
range from $L_G$ to $L$.  The estimated turbulent flame brush thickness is
proportional to $L$ and the linear relation between $S_T$ and $u'$ is obtained.

Another approach by Kerstein (1988b) estimates the turbulent flame brush in two
ways: 1) $\delta_T \sim \tau S_T$ and 2) $\delta_T \sim D/S_T$.  Here $\tau$ is
the consumption time for an eddy and the turbulent diffusivity $D$ is replaced
with $u' L$, giving
$$
      S_T^2 = u' L / \tau .
$$
The consumption time is estimated from the growth of flame area
by strain rate and Kerstein develops a simple derivation of Yakhot's result
obtained via renormalization techniques: $S_T/S_L = \exp[(u'/S_T)^2]$.  A
different estimate for the consumption time is based on an eddy spacing
proportional to the Taylor length scale $\lambda$ (see, G\"ulder, 1990) and the
laminar flame speed, yielding: $\tau \sim \lambda / S_L$.  This latter choice
leads to
$$
{S_T \over S_L} = \sqrt{ {u'\over S_L} {L\over \lambda} }
$$
and using the volume averaged energy dissiapation $\epsilon = 15 \nu
u'^2/\lambda^2 \sim u'^3/L$ we have
$$
{S_T \over S_L} = \sqrt{ {u'\over S_L} {Re_{\lambda}\over 15 } }  \eqno( 1)
$$
where $Re_{\lambda}$ is the Reynolds number based on the Taylor length scale.
In a latter section, we give results that appear to agree with the square-root
behavior given by Equation (1); these results are from numerical simulations of
passive flame propagation within Navier-Stokes turbulence and within defined
two-dimensional flows.

The above conceptual models of flame propagation through a turbulent flow leave
out the effects of volume expansion created by the chemical reaction.  However,
for the wrinkled flamelet regime, G\"ulder shows that a sampling of the
experimental data may follow the square-root power-law given above.  A
power-law with an exponent less than unity matches the character of the data
which has a linear relation when $u' \sim S_L$, but changes to an apparent
maximum for $S_T/S_L$ when $u' >> S_L$.  This changing relationship is also
called a ``bending" effect (see Peters, 1988 and Sivashinsky, 1988).  Volume
expansion, by damping the turbulence in the unburnt gas and thereby reducing
the flame wrinkling, could be one cause of this bending effect.  In the present
paper we propose that the spatial structure of turbulence could be another
cause of this bending effect.  All the computer simulations on which our
spectulation is based do not include heat release and so, we can not judge the
relative importance of these two possibilities for the bending effect.

The vital observation of the turbulence simulations is that the intense
vortical regions are tube-like in shape and that these tubes appear to be
finite in length.  Furthermore, the number density or the spacing of these
tubes is such that they are not space filling, see Figure 1.  The non-space
filling feature is consistent with Kolmogorov's newer theories regarding the
energy cascade which include the intermittency nature of turbulence (see
Gibson, 1991). Current direct simulations of turbulence have been done with
grid meshes of $\approx 100^3$, and so the observed turbulence structure will
need confirmation by larger calculations.  However, the current
three-dimensional turbulent flame propagation simulations do show how the
spatial structure of turbulence affects the turbulent flame speed.

\noindent
{\it Turbulence Structure Effect Upon Flame Propagation }

The effect of spatial structure is most obvious in a two-dimensional flow
composed of vortical eddys separated by regions which are non-swirling by
comparison, this yields a flame propagation which is composed of two types of
advancement.  One type is the flame propagation within the vortical eddy and
the other type is the flame motion between eddys.  Within an eddy, the swirling
flow will be characterized as having rigid body rotation (zero shear) near the
axis, a maximum swirl velocity at radius $R_m$ and a near zero swirl velocity
for radial locations beyond three or four times $R_m$, see Figure 2.  When the
swirl velocity $U_m$ is greater than $S_L$ then this eddy will wrinkle the
flame and when $U_m$ is much greater than $S_L$ this eddy will form pockets of
unburnt gas.  From time-dependent, two-dimensional simulations of flamelet
motion it becomes very obvious that when the flame contacts the eddy, contact
in the sense that some part of the flame surface is near the maximum swirl
radius, then the flame front begins to advance around the eddy circumference.
The rate of advancement is $U_m + S_L$.  After an elapsed time of approximately
half of a turn over time, the flame front will now be located a distance of $2
R_m$ from the flame location that existed before the contact.  When $U_m >>
S_L$ then the flame front has been significantly advanced during this fraction
of the eddy turn over time.  This is the eddy advancement of the flame.  The
non-eddy advancement corresponds to flame propagation in regions between eddys
and the flame duration in the non-swirling regions becomes the rate-limiting
step when the swirl velocity is much larger than the flame speed.

The fact that flow spatial structure may limit the maximum flame propagation
speed will be illustrated by numerical simulations of passive flame motion. The
two-dimensional eddy configuration to be presented is certainly a contrived
flow pattern.  However, it has been contrived to match a feature of turbulent
flow, namely, the intense swirling regions are not space filling.  The utility
of these two-dimensional simulations is the ease with which one can determine a
model of flame advancement which exhibits the rate-limiting behavior found in
the nonlinear simulations.  Examination of flamelet propagation within
three-dimensional Navier-Stokes turbulence reveals the same rate-limiting
behavior as found in the contrived flow.  Thus, flows in which the most intense
vortical motions are not space filling will exhibit a bending effect in the
relation of $S_T$ and $u'$.

The eddy structure shown in Figure 1 would be very simple if all the eddys had
the same swirl speed, swirl radius and tube length.  Only the arrangement in
three-dimensional space and the number density would need specification
(assuming that the eddy lifetime is longer than the flame passage time).  While
not proven at this time, we do suggest that the simple model is reasonable.
This concept was developed by trying to make the Burgers' vortex flow pattern
fit the strain rate behavior observed in turbulence simulations.  In order to
do so, the axial strain rate of the Burger's vortex is related to the
large-scale parameters $u'/L$ and the vortex circulation is a fixed multiple of
the kinematic viscosity.  This makes the eddy radius comparable to the Taylor
length scale.  Confirmation of this Burgers' vortex model must await further
simulations and the investigation of these simulations. The current value of
these suggestions is that they reduce the number of free parameters in the
structural description of turbulent flow and they show how the flow structure
determines the turbulent flame speed -- in the zero heat-release limit.

We first present the two-dimensional simulations which only include a single
length scale of eddys.  The functional form which describes these results also
gives a good representation of the three-dimensional Navier-Stokes results.  We
then make the assumption that these eddy flow patterns could be repeated on
larger length scales and so use the single-scale form as a recursion relation.

\vskip 24pt

\centerline{ Two-Dimensional, Single-Scale Eddys }

We treat a passive flame which has a propagation speed of $S_L$ and is moving
through a flow composed of swirling eddys.  The swirling motion of each eddy is
confined within a diameter $D$ and the eddys are spaced at distance $L$ and do
not overlap, $L>D$.  This collection of eddys creates a root-mean-square
velocity $u'$, and between the eddys the flow speed is zero in comparison with
$u'$, which we will refer to as the quiet zone.  When $u'>S_L$ the flame
advancement has two distinct paths: 1) propagation plus convection within an
eddy, and 2) propagation across the quiet zone to reach the next eddy.  The
effective velocity for the flame advancement is
$$
    S_f = { L \over t_1 + t_2 } \eqno(2)
$$
where $t_1 = l_1 / (S_L + u')$ and $t_2 = l_2 / S_L$.  The path length $l_1$ is
proportional to the eddy circumference when $u'>S_L$ and reduces to the eddy
diameter when $u'<S_L$.  The path $l_2$ is a representative distance of the
quiet zones.  The front speed is
$$
    S_f = { L \over { {\displaystyle l_1 \over\displaystyle S_L + u' } +
     {\displaystyle l_2 \over\displaystyle  S_L } } } \ \ \ \
   = { S_L + u' \over { {\displaystyle  l_1 \over\displaystyle L } +
{\displaystyle l_2 \over\displaystyle L } ( 1 + u'/S_L)  } }         \eqno(3)
$$
or, with $S_L$ = 1, in the form of
$$
    S_f = {(1 + u') \over (a + b u') } \eqno(4)
$$
with $a>1$ and $b<1$ for eddys that almost fill the available space.  When $u'$
increases without changing the number of eddys or the ratio $D/L$, then the
maximum front speed becomes $1/b = L/l_2$.  Thus, the quiet zones limit the
flame advancement and cause a {\it bending} effect in the relation $S_f$ vs
$u'$.

An illustration of this propagation is shown in Figure 3 in which two eddys
have been placed in a channel and the propagation dynamics have been solved
with the $G$ equation formulation
$$
{\partial G \over \partial t} + {\bf u}\cdot\nabla G = S_L |\nabla G| \eqno(5)
$$
(Kerstein {\it et al.,} 1988; Ashurst {\it et al.,} 1988).  Letting the total
flame surface equal the front advancement rate in this stationary system
$(<|\nabla G|> = S_f)$, then a fit of $<|\nabla G|>$ using Eq. (4) yields $a
\approx 1.05$ and $b \approx 0.3$, see Figure 4.

There is an apparent departure of the calculated response from the behavior
predicted by Eq. (4) when $u' \approx 2.5 S_L$.  This departure could be the
onset of unburned pockets within the eddy core: the rotation rate has convected
the flame clear around the eddy and further increases in rotation rate can not
further increase the flame area {\it and} with this complete wrap around there
is the formation of islands of unburnt.  Even with heat release, large eddy
rotation can create islands (see the two-dimensional simulations of Ashurst \&
McMurtry, 1989).  Flame pinching leading to pockets has been studied by Joulin
\& Sivashinsky (1991) in a non-swirling flow.  Pocket formation and its
transient effect could be included as a correction to Eq. (4) when $u' > 2S_L$.
The change in flame curvature seen in Figure 3 for $u'<S_L$ and for $u'>S_L$
indicates that the path length $l_2$ in Eq. (2,3) might have a change at $u'>2
S_L$.  From Figure 4, it appears that this path change should increase the
value of the $b$ coefficient and hence the maximum flame speed is reduced.  A
change in $l_2$ could be a change in the time to reach the maximum swirl radius
as the flame shape acquires a very large curvature.

Another two-dimensional eddy flow with the rate-limiting behavior is the
square-eddy flow given in Figure 5 of Ashurst \& Sivashinsky (1991): $u =
-\sqrt{2} u'\cos kx \sin ky; v = \sqrt{2} u'\sin kx \cos ky$.  The maximum
swirl velocity is at the edge of the eddy in this flow pattern, and so the
eddys are not separated by quiet zones.  However, a fit of the computed flame
speeds yields results similar to previous two-dimensional eddy results: the
coefficients for the square-eddys are $a = 1.07$ and $b = 0.11$.  The smaller
$b$ value in the square-eddy flow gives a larger maximum flame speed than is
possible for the same spatial average of $u'$ in the round-eddy configuration
given above.  In the square-eddy flow the quiet zone effect is created by the
stagnation region between four eddys (the flame points are in the
stagnation regions in sequence 4 of Figure 5 cited above).  So, we may conclude
that two-dimensional swirling motions will inherently have quiet regions which
limit the maximum flame speed.  The quiet-zone effect on flame propagation may
also occur in three-dimensional turbulence due to the non-space filling nature
of the turbulence structure.

\vskip 24pt

\centerline{ Three-Dimensional Flame Propagation }

Three-dimensional Navier-Stokes turbulence simulations have been combined with
the $G$ equation to determine the flame speed dependence upon $u'/S_L$.  In
these simulations the turbulent kinetic energy is maintained at a constant
value by weak forcing of the large-scale strain rate.  The flame curvature
distribution given by these simulations, at $S_L = 2 u'$, has good agreement
with experimentally determined curvature, see Shepherd \& Ashurst (1992).  The
flame speed results given in Figure 5 were obtained in a periodic cube with
32$^3$ grid-cells, the value of $u'$ is unity and the kinematic viscosity is
$\nu = u' L / 500$ where $L$ is the cube edge length, also equal to unity.  The
measured Taylor length scale is $\lambda \sim L/8$, resulting in $Re_\lambda
\sim 60$. The laminar flame speed $S_L$ was varied while the values of $u'$ and
$\nu$ were fixed.  Each turbulent flame speed value is a time average over the
same turbulence simulation, that is each initialization of the $G$ equation has
the same flow solution, the averaging time-period is $32 L/u'$, about 28,000
time-steps.  A fit of the three-dimensional, Navier-Stokes flame speed results
using Eq. (4) yields $a = 0.9021$ and $b = 0.1585$, values which are similar to
the two-dimensional eddy results.

Examination of the computed turbulence reveals that the intense vortical eddys
are separated by quiet zones and so, flame speed behavior in swirling flows
may not depend upon flow dimensionally.  Details of the turbulence
structure are given by She {\it et al.} (1990).  A noteworthy feature for
modeling flame propagation is that a random flow, but with the Navier-Stokes
energy spectrum, does not have the same vortical structure as the turbulent
solution (see pictures in the {\it Nature} article).

An analysis of preliminary numerical flame speed results indicated that a
square-root behavior might be the best fit, see Wirth \& Peters (1992).
Therefore, in Figure 6, the current $32^3$ results are plotted versus
$\sqrt{u'/S_L}$.  The slope of the straight line is 1.32.  We have used this
square-root behavior in order to determine the $a, b$ coefficients in a simple
manner by expressing Eq. (4) in terms of $\sqrt{u'/S_L}$.  Let $x^2 = u'$ and
then Eq. (4) is
$$
   S_f(x) = {1 + x^2 \over a + b x^2 } \eqno(6)
$$
and the slope of this function in terms of $x$ is $2 x (a-b)/(a+b x^2)^2$.
For $u' = S_L$, denote the slope as $m_1$ and the flame speed value as $S_f(1)$
then
$$
  (a - b) = 2 m_1 / S^2_f(1) \ \ \ \ \ \ \ \ (a + b) = 2 / S_f(1) . \eqno(7)
$$
This procedure has been used to determine the coefficients of the
two-dimensional eddy flows given previously.  Examination of the dependence of
the $a,b$ coefficients, given by Eq. (7), upon the slope $m_1$ and the flame
speed $S_f(1)$ shows that the slope $m_1$ has a stronger influence on $b$ than
on $a$, and the opposite behavior occurs for $S_f(1)$.

The square-root behavior seen in Figure 6 is consistent with the turbulence
model given in Eq. (1), except for small values of $u'$ where the numerical
results do not appear to converge towards the value of $S_T = S_L$.  In
application of the model given in Eq. (1), G\"ulder replaces the left-hand side
with $(S_T/S_L) - 1$ so that as $u' \rightarrow 0$, then $S_T \rightarrow S_L$.
A recent analysis of weak, turbulent flame propagation indicates that neither
the square-root nor the quadratic behavior will occur in a random field.
Instead, a 4/3 power-law is realized in random fields when $u'<<S_L$ (Kerstein
\& Ashurst, 1992).   Therefore, the dashed line connecting to the laminar flame
speed value in Figure 6 is obtained by passing a 4/3 power relation through the
first data point.  The resulting coefficient is of order unity (1.43) and the
smoothness of the transition is agreeable.

\noindent
{\it Flamelet Reynolds Number Dependence }

G\"ulder (1990) has shown that experimental data for the wrinkled flamelet
regime does appear to follow the square-root behavior given in Eq. (1).  We
have used the Taylor based Reynolds number as that quantity can be directly
calculated in the simulations.  The computational domain is not large enough to
obtain a reliable measure of the integral scale, and the grid size is too
coarse for the Kolmogorov scale, however the current simulations do have good
agreement with strain-rate distributions obtained in larger-grid simulations.
Therefore, we determine the Taylor Reynolds number and use $\sqrt{Re_L} =
Re_\lambda = Re_\eta^2$ to relate the different scales.  From Eq. (1), the
estimated slope for the $32^3$ results would be $\sqrt{ 60 / 15} = 2$ whereas
the numerical simulations yield 1.32.  A few simulations have been done with a
$64^3$ grid and, with this better resolution, the Taylor length scale
changes to $\sim L/9$, giving an estimated slope, from Eq. (1), as
$\sqrt{55 /15} = 1.91$.  Connecting the first and second points obtained with
$64^3$ yields a slope of 1.91.  The uncertainity in the numerical values is at
least $\pm 5\%$.  Even larger grid simulations have not been done, but with the
latter comparison, it appears that the consumption time estimate based on the
Taylor length scale and the laminar flame speed yield a turbulent flame brush
model which agrees with the numerical results.

Simulations with other viscosity values would allow comparison with the
Reynolds number scaling.  The current conditions are at the border of the
flamelet regime given by G\"ulder, the bounds being $\eta > 1.5 (\nu/S_L)$ and
$Re_L < 3200$ or $Re_\lambda < 57$ (the reason for the Reynolds limit is not
given, nor is it consistent with other flame diagrams, {\it cf.} Peters, 1988).
Anyway, from G\"ulder's view the current simulations are at the upper range of
Reynolds number.  The curent simulations may also be the minimum $Re_\lambda$
for turbulent energy dissipation to be independent of Reynolds number.
For example, in grid turbulence, Sreenivasan (1984) shows that the turbulent
energy dissipation, normalized by $u'^3/L$, depends upon Reynolds number when
$Re_\lambda < 50$.  This dissipation Reynolds dependence could be evidence that
the vortical structure is different in this regime, and so flame propagation
could also have a different Reynolds number behavior.  Experimental
determination of turbulent flame speed could also have a dependence upon how
the turbulence is generated.  Sreenivasan presents results in which changes in
grid geometry, from square holes to parallel rods, yields factors of two in the
downstream dissipation value.  Such grid effects might explain the larger
coeficient in the flame results of Liu \& Lenze cited by G\"ulder.

\vskip 24 pt

\centerline{ Multiple-Scale Eddys }

So far we have considered only a single scale of eddys with quiet zones between
eddys.  The three-dimensional Navier-Stokes simulations are essentially a
single-scale vortical structure, probably due to the small grid size.  In the
spirit that nature repeats patterns at different length scales, we conjecture
what would be the effect upon the flame speed if there were a hierarchy of eddy
scales.  However, we wish to think of the smaller eddys as occurring between
the larger eddys, and not within them.  We exclude the small eddys from the
region occupied by the larger eddys because the differential rotation caused by
the intense vorticity removes velocity fluctuations in the large swirl regions.
At larger radial distances from an eddy, the shearing motion becomes the
negative of the swirling motion and irrotational exists.  In this outer region
of the large eddy, a small eddy would be convected but not distorted.
Therefore, if we let the in-between zones of large eddys be occupied by similar
eddys on a smaller length scale, then the time $t_2$ in Eq. (2) should be
changed to
$$
   t_2 = l_2 / S_{f_2}  \eqno(8)
$$
where $S_{f_2}$ replaces $S_L$ in what was previously assumed to be a
quiet zone and so Eq. (3) changes to
$$
  S_{f_1}   = { S_L + u' \over { {\displaystyle l_1 \over\displaystyle  L } +
{\displaystyle l_2 \over\displaystyle L }  (S_L + u') / S_{f_2} } } . \eqno(9)
$$
It appears that by using functional iteration we may include the effects of
eddys at smaller, or larger, length scales.  We change Eq. (4) to
$$
  S_{f_n} = { 1 + u' \over { (a - b) + b ( 1 + u') / S_{f_{n-1}} } }  \eqno(10)
$$
where $S_{f_n}$ is the effective flame speed for eddys at spacing $l_n
(>l_{n-1} )$ and at the smallest possible eddy length scale $l_1$, $S_{f_0}$ is
$S_L (=1)$ and we have Eq. (4).  A flow with $p$ levels of eddys will have a
maximum speed of $S_L / b^p$ and so the bending effect will still appear, but
at a larger value of $u'$ than a flow with $p-1$ levels.  In this eddy model
the propagation within an eddy is at $S_L + u'$ and only between the eddys is
there an enhanced propagation of $S_{f_{n-1}}$, which is different than
Sivashinsky's model (1988) where each length scale $l_n$ distorts a flame that
travels with an apparent speed of $S_{f_{n-1}}$.  The eddy pattern proposed
here is that of smaller eddys {\it between} larger eddys rather than eddys {\it
within} larger eddys.

Using this functional iteration procedure we can take the numerical simulation
as the smallest eddy size and expand the results to a larger volume with eddys
active on many length scales.  An example of this procedure is given in the
next section.

\vskip 24pt

\centerline{ Comparison with Liquid Flames }

Ronney and co-workers (Shy {\it et al.}, 1992) have used a chemically
reacting, propagating front in a liquid to simulate thin premixed flames which
have many of the popular theoretical assumptions: density, $S_L$ and transport
coefficients are all constant.  They obtain flame speed versus turbulence
intensity in two different configurations: 1) Taylor-Couette and 2)
capillary-wave flow.  In these liquid flames they can reach values of $u' =
1000 S_L$ and more. Over a wide range of $u'$ they find the Taylor-Couette
results are reasonably described by $S_T = \exp( u'/S_T)$, with $S_L = 1$.

We use this observation of their experimental results to determine values in
the functional iteration procedure.  Using Eq. (10) in a repeated fashion, we
assume that the introduction of the next level of eddys should produce a flame
speed $S_{f_n}$ which does not exceed the value given by the relation $S_T \ln
S_T = u'$.

In Figure 7, the Navier-Stokes $a,b$ values from above are maintained constant
as new levels of eddys are introduced.  In order to achieve no overshoot, as
described above, we find that the turbulence parameter $u'$ depends on the eddy
level, in fact decreasing as we include larger length scales.  Rather than
kinetic energy of the flow, it is closer to the eddy flow pattern to think of
$u'$ as representating the eddy vorticity times the swirl radius -- perhaps the
Burgon structure (Ashurst, 1992).  In a Burgers' vortex the vorticity has a
Gaussian radial dependence and the turnover time at the maximum swirl radius
$r_m$ is $t_u \approx 1/s$ where $s$ is the axial strain rate (constant $s$
gives a steady, viscous flow). The maximum-swirl radius scales as $r_m \propto
\sqrt{4\nu/s}$ and with $u' \propto r_m /t_u$ then $u' \propto \sqrt{4\nu s}$.
Hence, the parameter $u'$ should increase as the length scale decreases because
smaller scales have larger strain rates.

The introduction of one eddy level larger than the Navier-Stokes simulation,
Figure 7a, with $u'_2/u'_1 = 0.42$, provides agreement for $u'<30$.  In Figure
7b, with $u'_3/u'_2 = 0.62$, the agreement goes beyond 200 in $u'$.  In Figure
7c, with $u'_4/u'_3 = 0.70$, there is agreement up to $u' = 1000$.  While these
assumed ratios of swirl intensity are not to be taken too seriously, it is
worth noting that only a few eddy levels are needed to give the appearence of a
turbulent flame speed that is similar to experimental expressions.  {\it If}
this eddy structure is close to the actual physical motion, then a turbulent
flow with only two or three eddy levels would appear to have $S_T \sim u'$ when
$u' > S_L$.

\vskip 24pt

\centerline{ Features of the Eddy Model }

This flame propagation model depends on the fluid vorticity structure, the
eddys.  From experiments and turbulence simulations, we know that two eddys can
not be too close to each other if they are to remain as distinct eddys.
Between a pair of eddys there usually is a local stagnate region, and it is
this quiet zone around the eddy which is the rate limiting step in flame
propagation when the eddy swirl speed becomes larger than the flame
speed.  However, if it is the nature of turbulence to create similar flow
patterns at different length scales, then in the quiet zone there can be
another pair of eddys, but much smaller in size. These small-scale eddys
enhance the quiet zone flame propagation and so the rate limiting step of the
larger scale has been changed by the physical location of the small scales.
Note that this model has the smaller eddys between the larger ones, not within
them.  This eddy pattern can be repeated down to the level where the molecular
structure causes departure from the continuum solution or to the level where
the energy flux is not strong enough to create distinct eddys.  The Burgers'
vortex is a good candidate for the eddy structure, this viscous, dissipative
flow has a steady solution when the axial strain rate $s$ is constant.  The
vortex vorticity times swirl radius may scale as $\sqrt{\nu s}$, which we
assume is the relevant parameter $u'$ in the flame speed relation.  By
adjusting $u'_n$ for each level of eddys at length $l_n$, we have shown that it
is possible to approximate a flame speed relation like $S_T \ln S_T = u'$.  We
have not specified the ratio $l_n / l_{n+1}$.  Further large-scale simulations
of turbulence, at constant energy, should allow determination of a multi-level
eddy structure.  We wonder if these eddy levels will exhibit the universality
seen in other dynamical systems.

\vskip 24pt

\centerline{ Comments }

We can relate the proposed eddy structure to the coherent structures that occur
in mixing layers, both two and three dimensional (Ashurst, 1979 and Ashurst \&
Meiburg, 1988).  These moving structures produce a smooth time-average Eulerian
velocity which does not resemble the Lagrangian flow pattern around any
structure.  The important point is that flame motion occurs in the Lagrangian
frame.  The difficulty is that the usual experimental determination is an
Eulerian measure.  Grasping how different these two viewpoints can be, when
molecular mixing determines the phenomena being considered, requires unsteady
information, either from simulations or from global diagnostics, such as laser
sheet images.

In turbulent combustion the overall relation of $S_T$ versus $u'$ could depend
on how the turbulence changes as the magnitude of $u'$ is increased: do more
levels of eddys appear in a smooth fashion or as discrete jumps?  Is the
behavior universal in the Feigenbaum sense, that is can the amplitudes and
levels be predicted?  There has been abundant work on chaos in systems
with linear diffusion, here we have nonlinear diffusion, the $G$ equation,
in simple structured flows such as the eddy-quiet-zone pattern -- does this
lead to similar chaotic behavior?

Most Fourier-based concepts of turbulent flow ignore the phase angle
information.  She {\it et al.} (1990) give a graphic illustration of the
difference in flow pattern between two systems which have the same energy
sprectrum, but one flow is a Navier-Stokes solution and the other has random
phase -- the intense vorticity structure, the eddys, do not look at all alike.
Flame propagation, we suspect, is very dependent on the actual flow pattern
rather than just the energy spectrum.

The iteration equation given by this simple eddy model may allow direct
numerical simulations to be iterated to volumes of engineering interest.  There
appears to be an analogy with progress in molecular dynamics: computer
simulation of the hard-sphere fluid led to a polynominal description of
pressure versus density.  From this base equation of state it is possible to
analytically include the effects of an attractive power and the resulting
desired equation of state.  Similar success was obtained for momentum and
thermal transport: computer simulation of soft-sphere viscosity and thermal
conductivity gives the density dependence which can be added to the temperature
dependence given by kinetic theory (Ashurst \& Hoover, 1975).

Therefore, simulation of flame propagation within Navier-Stokes turbulence
structure may yield the nonlinear relation of flame speed and turbulence
intensity.  Addition of larger eddys, outside the range of computer memory, may
be accounted for by specification of the coefficients $a_n, b_n$ (assumed
invariant in the current work) and $u'_n$, where $n$ is the eddy level.  Which
scaling laws and/or fluid mechanical physics should guide the selection of
these coefficients?  One proposal is the Gurvich-Yaglom model of turbulent
energy dissipation (Kolmogorov's third hypothesis), which relates the mean and
variance of dissipation to observational volume size, and so the coefficients
obtained in a direct simulation could be used to estimate the dissipation
statistics at larger sizes (Kerstein \& Ashurst, 1984).  A second proposal
would exploit the observed strain rate behavior dependence upon strain
magnitude: the shape of the strain rate tensor tends to a triangular symmetric
probability distribution for the intermediate strain rate $\beta$ at low strain
(Ashurst {\it et al.}, 1987).  Hence, the most probable shape is that of plane
strain for the larger length scales.  So, small-scale direct simulations of
flame propagation, combined with an iteration procedure which incorporates the
fluid structure, may determine the turbulent flame speed for an arbitrarily
large volume of turbulence (arbitrarily large in the galactic sense, {\it cf.}
Gibson, 1991).

\vfill\eject

\centerline{ Acknowledgement }

Discussions with Alan Kerstein, Norbert Peters and Paul Ronney have been
helpful.  The insight gained by viewing Figure 1 from the work of She, Jackson
\& Orszag has been a crucial aspect of this work; permission to reproduce the
figure is gratefully acknowledge.  This work supported by the United States
Department of Energy through the Office of Basic Energy Sciences, Division of
Chemical Sciences.

\vskip 24pt

\centerline{ References }

\parindent 0 pt

Ashurst, Wm. T. \& Hoover, W. G. (1975). Dense-fluid shear viscosity via
nonequilibrium molecular dynamics.  {\it Phys. Rev. A} {\bf 11}, 658.

Ashurst, Wm. T. (1979). Numerical Simulation of Turbulent Mixing Layers via
Vortex Dynamics.  {\it Turbulent Shear Flows I}, F. Durst, {\it et al.}, Eds.
(Springer-Verlag) 402.

Ashurst, Wm. T., Kerstein, A. R., Kerr, R. M. \& Gibson, C. H. (1987).
Alignment of vorticity and scalar gradient with strain rate in simulated
Navier-Stokes turbulence.  {\it Phys. Fluids} {\bf 30}, 2343.

Ashurst, Wm. T. \&  Meiburg, E. (1988). Three-Dimensional shear layers via
vortex dynamics.  {\it J. Fluid Mech.} {\bf 189}, 87.

Ashurst, Wm. T., Sivashinsky, G. I. \& Yakhot, V. (1988). Flame Front
Propagation in Nonsteady Hydrodynamics Fields. {\it Comb. Sci. \& Tech. }
{\bf 62}, 273.

Ashurst, Wm. T. \& Mc Murtry, P. A. (1989). Flame Generation of Vorticity:
Vortex Dipoles from Monopoles. {\it Comb. Sci. \& Tech. } {\bf 66}, 17.

Ashurst, Wm. T. \& Sivashinsky, G. I. (1991). On Flame Propagation Through
Periodic Flow Fields. {\it Comb. Sci. \& Tech. } {\bf 80}, 159.

Ashurst, Wm. T. (1992). Is Turbulence a Collection of Burgers' Vortices?
Submitted to {\it Phys. Fluids A}.

Clavin, P. \& Siggia, E. D. (1991). Turbulent Premixed Flames and Sound
Generation.  {\it Comb. Sci. \& Tech. } {\bf 78}, 147.

Gibson, C. H. (1991). Kolmogorov similarity hypotheses for scalar fields:
sampling intermittent mixing in the ocean and galaxy. {\it Proc. R. Soc. Lond.
A} {\bf 434}, 149.

G\"ulder, \"O. L. (1991).  Turbulent Premixed Flame Propagation Models for
Different Combustion Regimes.  {\it Twenty-Third Symposium (International) on
Combustion/The Combustion Institute}, 743.

Joulin, G. \& Sivashinsky, G. I. (1991). Pockets in Premixed Flame and
Combustion Rate. {\it Comb. Sci. \& Tech. } {\bf 77}, 329.

Kerstein, A. R. (1988a).  Fractal Dimension of Turbulent Premixed Flames.
{\it Comb. Sci. \& Tech. } {\bf 60}, 441.

Kerstein, A. R. (1988b).  Simple Derivation of Yakhot's Turbulent Premixed
Flamespeed Formula. {\it Comb. Sci. \& Tech. } {\bf 60}, 163.

Kerstein, A. R. \& Ashurst, Wm. T. (1984). Lognormality of gradients of
diffusive scalars in homogeneous, two-dimensional mixing systems.
{\it Phys. Fluids} {\bf 27}, 2819.

Kerstein, A. R., Ashurst, Wm. T. \& Williams, F. A. (1988). Field Equation for
Interface Propagation in an Unsteady Homogeneous Flow Field. {\it Phys. Rev. A}
{\bf 37}, 2728.

Kerstein, A. R. \& Ashurst, Wm. T. (1992). Propagation Rate of Growing
Interfaces in Stirred Fluids. {\it Phys. Rev. Lett.} {\bf 68}, 934.

Peters, N. (1988). Laminar Flamelet Concepts in Turbulent Combustion.
{\it Twenty-First Symposium (International) on Combustion/The Combustion
Institute}, 1231.

She, Z.-S., Jackson, E. \& Orszag, S. A. (1990). Intermittent vortex structures
in homogeneous isotropic turbulence. {\it Nature} {\bf 344}, 226.

She, Z.-S., Jackson, E. \& Orszag, S. A. (1991). Structure and dynamics of
homogeneous turbulence: models and simulations.  {\it Proc. R. Soc. Lond. A}
(1991) {\bf 434}, 101.

Shepherd, I. G. \& Ashurst, Wm. T. (1992). Flame Front Geometry in Premixed
Turbulent Flames. {\it Twenty-Fourth Symposium (International) on
Combustion/The Combustion Institute}, 485.

Shy, S. S., Ronney, P. D., Buckley, S. G. \& Yakhot, V. (1992). Experimental
Simulation of Premixed Turbulent Combustion Using Aqueous Autocatalytic
Reactions.  {\it Twenty-Fourth Symposium (International) on Combustion/The
Combustion Institute}.

Sivashinsky, G. I. (1988). Cascade-Renormalization Theory of Turbulent Flame
Speed. {\it Comb. Sci. \& Tech. } {\bf 62}, 77.

Sreenivasan, K. R. (1984). On the scaling of the turbulence energy dissipation
rate. {\it Phys. Fluids}, {\bf 27}, 1048.

Wirth, M. \& Peters, N. (1992). Turbulent Premixed Combustion: A Flamelet
Formulation and Spectral Analysis in Theory and IC-Engine Experiments.
{\it Twenty-Fourth Symposium (International) on Combustion/The Combustion
Institute}.

\vfill\eject
\centerline{ Captions }

Figure 1.  Computed Navier-Stokes turbulence reveals the tube-like structure of
intense vorticity (dark lines) compared to the lack of structure for moderate
levels of vorticity (grey lines).  The grid size is $96^3$ with $Re_\lambda
\sim 77$, figure from She, Jackson \& Orszag (1991).

\vskip 6pt

Figure 2.  Vortical eddy effect upon premixed flame shape.  The character of
a viscous vortex is a maximum swirl velocity ($U_m$ located at $R_m$) which
creates a flame tip in the advancing flame.

\vskip 6pt

Figure 3.  Front propagation through two-dimensional eddys is shown by ten
contours of $G$ (any constant $G$ surface represents a flame in this passive
model). Propagation is from right to left and each eddy is confined within a
radius of 1/2 with the maximum swirl speed at a radius of 1/6.  Notice the
flame tip formation at the radial location of maximum swirl.  This eddy shape
resembles a viscous vortex and is described in Kerstein \& Ashurst (1992).  The
swirl rates are $u' = 0.6 S_L$ in a) and 2.0 in b) with grid sizes of 64 by
128, and 128 by 256, respectively.  Note the incipient unburned pockets in the
eddy core at large $u'$.

\vskip 6pt

Figure 4.  Front advancement rate versus eddy swirl speed $u'/S_L$ for the
two-dimensional round-eddy configuration shown in Fig. 3.  The numerical
results are $S_T/S_L = <|\nabla G|>$ (filled circles) and the model
approximation, Eq. (4), is shown by the solid line.

\vskip 6pt

Figure 5.  Calculated turbulent flame speed $S_T/S_L = <|\nabla G|>$ in
three-dimensional Navier-Stokes turbulence (filled circles) also compares well
with the swirling model approximation, Eq. (4), shown by solid line.

\vskip 6pt

Figure 6.  The calculated Navier-Stokes turbulent flame speed results given in
Figure 5 appear to have a square-root behavior (filled circles: $32^3$ grid,
triangles: $64^3$ grid).  The straight line is determined from the slope given
by the first and fifth calculated points using the $32^3$ grid.  This slope (=
1.32) combined with the flame speed value at $u'=S_L$ gives $a, b$ values of
0.9021 and 0.1585. The dashed line from $S_T = S_L$ connects the first point
with a 4/3 power as found in a random eddy flow at $u'<<S_L$ (Kerstein \&
Ashurst, 1992).

\vskip 6pt

Figure 7.  Functional iteration with Eq. (10) can approximate the relation $S_T
\ln S_T = u'$, the latter is observed in liquid flame experiments.  $S_{f_1}$
is the Navier-Stokes results given in Figure 6 and $S_{f_2}, S_{f_3}$ and
$S_{f_4}$ are obtained by changing $u'_n$, but not the values of $a,b$.

\end